\documentclass{svjour3}                     % onecolumn (standard format)
\smartqed  % flush right qed marks, e.g. at end of proof
\usepackage{graphicx}
 \journalname{Hyperfine Interactions}
\begin{document}

\title{Radium ion: A possible candidate for measuring atomic parity violation}


\author{P.~Mandal\and A.~Sen\and M.~Mukherjee}


\institute{P. Mandal \at Raman Center for Atomic Molecular and Optical Sciences\\
               Indian Association for the Cultivation of Science\\
                2A \& 2B Raja S. C. Mullick Road, Kolkata 700032, India\\
              Tel.: +91-33-24734971\\
              \email{drupm@iacs.res.in}           %  \\
          % \and
}

\date{Received: date / Accepted: date}

\maketitle

\begin{abstract}
Single trapped and laser cooled Radium ion as a possible candidate for measuring the parity violation induced frequency shift has been discussed here. Even though the technique to be used is similar to that proposed by Fortson \cite{Ref1}, Radium has its own advantages and disadvantages. The most attractive part of Radium ion as compared to that of Barium ion is its mass which comes along with added complexity of instability as well as other issues which are discussed here.
\end{abstract}
\section{Introduction}
\label{intro}
\begin{table}[width=textwidth]
\centering
\caption{Different techniques for PNC measurement and their advantages and disadvantages. $^{\ast}$Proposed techniques, yet to demonstrate.}
\label{tab1}       % Give a unique label
\begin{tabular}{lll}
\hline\noalign{\smallskip}
Techniques & Advantages & Disadvantages  \\
\noalign{\smallskip}\hline\noalign{\smallskip}
Optical rotation in & No electric and magnetic fields & Unavoidable systematic effects, \\
atomic vapor \cite{Ref6,Ref7,Ref8} & are involved, no frequency & poor signal to noise ratio at zero \\
 & measurements &  crossing in the dispersion curve \\
\hline
Stark interference & Measurement procedure is & Measured transitions are \\
in atomic vapor \cite{Ref14} & relatively simple & Doppler broadened \\
\hline
Stark interference & Doppler broadening is reduced, & Limited by volume and time \\
in atomic beams & signal to noise ratio is larger & of interaction, coherence \\
\cite{Ref4,Ref5,Ref12} & due to large no. of atoms & time is short due to collision \\
\hline
$^{\ast}$Light-shift in single & Absence of Doppler broadening, & Accurate determination of the  \\
trapped and laser & tractable systematic, long & electric field of the light at the \\
cooled ion \cite{Ref1,Ref11} & coherence time, large signal to & position of the ion in the trap \\
 & noise ratio & \\
 \hline
$^{\ast}$Stark interference & Large signal to noise ratio & Less systematic from collision \\
with small number &  & broadening \\
of atoms \cite{Ref10} & & \\
\noalign{\smallskip}\hline
\end{tabular}
\end{table}
Weak interaction between atomic electron and the nucleus through the exchange of Z$_{0}$ boson leads to parity violation in atomic systems \cite{Ref2}. Atomic parity violation (APV) has become a subject of keen interest as it has the potential to test the Standard Model (SM) of particle Physics and to search for new Physics beyond it \cite{Ref3}. Several experiments have been performed over the last three decades on some heavy elements like Cs \cite{Ref4,Ref5}, Pb \cite{Ref6}, Tl \cite{Ref7}, Bi \cite{Ref8} \emph{etc}. There are also some proposals with promising prospects on elements like Yb \cite{Ref9}, Fr \cite{Ref10} and atomic ions like Ba$^{+}$ \cite{Ref1} and Ra$^{+}$ \cite{Ref11}. One of the most promising candidates is Yb whose parity non-conserving (PNC) amplitude $E1_{PNC}$ so far the largest. This point has also been verified experimentally \cite{Ref12} but the experimental precision needs to be improved in order to compete with the present bench mark value of Cs PNC experiment \cite{Ref4}. The experiment on Cs with an accuracy of $0.35 \%$, has successfully explained the SM of particle Physics \cite{Ref4}. Higher precision ($0.1 \%$) is required to search for new Physics beyond SM \cite{Ref13}. The physical parameter that one seeks by combining these experiments and theory is the PNC transition amplitude $E1_{PNC}$. In Table~\ref{tab1} presently available techniques have been mentioned along with their advantages and respective challenges. A single trapped and laser cooled ion is free from unknown perturbations and it has long coherence time. Systematic uncertainties are easily tractable and therefore, the system is more favored for such experiment \cite{Ref1} even though this has not yet been experimentally demonstrated.

\section{Experimental Idea}
\label{sec:1}
\begin{figure}
\centering
  \includegraphics[width=0.85\textwidth, angle=270]{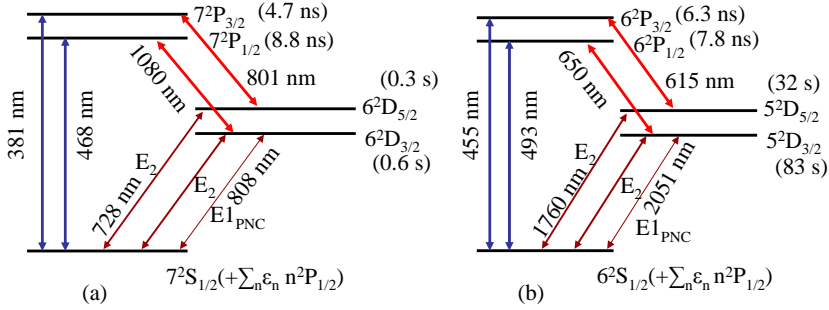}
\vspace{-5.5cm}
\caption{Relevant energy levels of (a) Ra$^{+}$ and (b) Ba$^{+}$}
\label{fig1}       % Give a unique label
\end{figure}

Single ion trapping and laser cooling are routinely done in radio frequency Paul traps \cite{Ref15}. The possibility of APV experiment based on such a system was first put forward by Fortson \cite{Ref1}. The overall idea has been reviewed here in brief focusing Ra$^{+}$ as a possible candidate. In Fig.~\ref{fig1} the relevant energy levels of singly charged Radium (Ra$^{+}$) and Barium (Ba$^{+}$) have been shown. After confining Radium ion in an RF Paul trap, it can be laser cooled by exciting the $S_{1/2} - P_{1/2}$ transition at $468$ nm. A repumping laser at $1080$ nm is necessary to bring the ion back to the cooling cycle from the metastable $6D_{3/2}$ state. Atomic parity violation leads to mixing of different parity states with the ground $7S_{1/2}$ state. Thus the ground state has a small contribution from $7P_{1/2}$ state resulting in a non-zero probability of dipole transition between $7S_{1/2}$ and $6D_{3/2}$ states which is normally a forbidden electric dipole transition.

\begin{figure*}
\centering
  \includegraphics[width=0.75\textwidth]{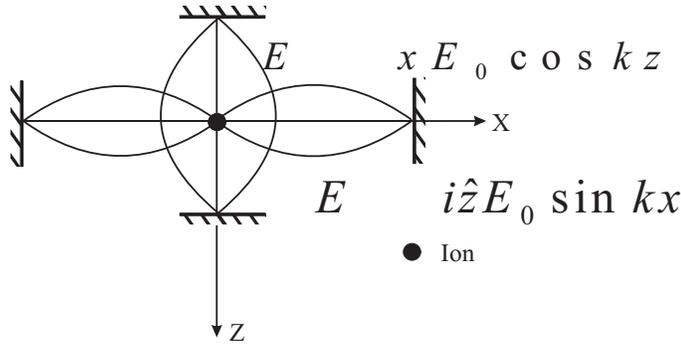}
\caption{A schematic diagram of a possible standing wave laser configuration to produce detectable APV light-shift on a single trapped Ra ion. The laser wave anti-node along z-axis generates a APV light-shift while the x-axis node produces a light-shift due to a quadrupole transition amplitude. These two light-shifts interfere to give the APV signal.}
\label{fig2}       % Give a unique label
\end{figure*}

A transitional dipole interacts with the electric field while a quadrupole interacts with the field gradient. In an experimental setup as shown in Fig.~\ref{fig2}, it is possible to induce both a dipole transition (due to APV) as well as a quadrupole transition between $7S_{1/2}$ and $6D_{3/2}$ states. The interference term of these two leads to a measurable frequency change of the Larmor frequency between the ground state Zeeman sublevels in presence, as compared to, in absence of the laser fields. One of the suitable laser field configurations that produce the needed APV frequency shift is
\begin{equation}
E'=\hat{x}E_{0}'\cos kz
\end{equation} \&
\begin{equation}
E''=i\hat{z}E_{0}''\sin kx,
\end{equation}
where $E'_{0}$ and $E''_{0}$ are the electric field amplitudes of the two lasers. An ion placed at the antinode of $E'$ field will suffer PNC induced electric dipole light-shift while the ion placed at the node of $E''$ field, will show electric quadrupole light-shift. The quadrupole light-shifts of the Zeeman sublevels in the ground state due to the $E''$ field are of the same magnitude and direction. Therefore, $E''$ field will not lead to any change of the ground state Larmor frequency defined by the energy difference between the Zeeman sublevels of the ground state. On the contrary, the shifts due to $E'$ field will increase the Larmor frequency. This change in Larmor frequency is proportional to the magnitude of the $E'$ field.

In the experiment one measures the Larmor frequency with and without these laser fields. The difference of these two frequencies therefore, gives directly the APV light-shift $\Delta\omega^{PNC}_{m}$ which can be expressed as \cite{Ref1}
\begin{equation}
\Delta\omega^{PNC}_{m}\cong -Re\sum_{m'}(\Omega^{PNC\ast}_{m'm}\Omega^{quad}_{m'm}/\Omega^{quad}_{m}),
\end{equation}
where $\Omega^{PNC\ast}_{m'm}$ and $\Omega^{quad}_{m'm}$ are the Rabi frequencies for PNC and quadrupole induced transitions which are respectively proportional to the electric field amplitude and field gradient of the standing wave lasers, $(\Omega^{quad}_{m})^{2}\equiv\sum_{m'}|\Omega^{quad}_{m'm}|^{2}$; $m$, $m'$ are the Zeeman sublevels of $S_{1/2}$ and $D_{3/2}$ states respectively. The distinguished advantage of this technique is that measurement of $\Delta\omega^{PNC}_{m}$ is free from any fluctuation in the laser frequency and other sources of quadrupole shift ($\Delta\omega_{Q}$). The statistical uncertainty in the measurement of $E1_{PNC}$ is given by
\begin{equation}
\delta E1_{PNC}=\frac{\hbar}{E'_{0}f\sqrt{Nt\tau}}
\end{equation}
where $f$ is an efficiency factor that depends on experimental conditions, $N$ and $\tau$ are the number of ions and coherence time respectively and $t$ is the time of observation. Though $N = 1$ in this experiment, longer coherence time improves the uncertainty in the measurement.
Accurate determination of $E1_{PNC}$ from measured $\Delta\omega^{PNC}_{m}$ depends on precise determination of the electric fields $E'$ and $E''$ at the position of the ion in the trap which is a challenge of this experiment and is a major source of systematic uncertainty. Also placing the ion at the antinode of $E'$ field or node of $E''$ field is a difficult task. Since PNC induced light-shift ($\Delta\omega^{PNC}_{m}$) is very small ($2\pi \times 5.3$ cycles/s for $E_{0}' = 2\times10^{6}$ V/m for Ra$^{+}$), little fluctuations of the magnetic field resolving Zeeman sublevels, will worsen the accuracy of the result.

However, in the past decades several experimental techniques have been developed by which the above problems can be solved. One can manipulate the ion position with respect to the standing wave. Recently, a technique has been reported \cite{Ref16} by which the nodal point or the nodal line of the trap can be shifted upto few micrometers. Thus the ion can be placed on the geometrical line of the standing wave. Systematic uncertainty originating from inaccurate positioning of the ion with respect to the electric field of the laser has been determined in the following. Since the ion is cooled to Lamb - Dicke regime, its motion is confined within its de Broglie wavelength ($\lambda_{de}$), typically $50$ nm for Ra$^{+}$ and Ba$^{+}$. The uncertainty in $E'$ and $E''$ fields for antinodal and nodal positions respectively are
\begin{equation}
\frac{\Delta E'}{E'_{0}}\approx (1-\cos k\lambda_{de})
\end{equation} and
\begin{equation}
\frac{\Delta E''}{E''_{0}}\approx \sin k\lambda_{de}
\end{equation}
These systematic uncertainties tabulated in Table~\ref{tab2}, have been estimated from the Lamb - Dicke parameter. There are also some techniques for controlling magnetic field fluctuations. Recent improvement of high precision RF spectroscopic techniques \cite{Ref11,Ref17,Ref18,Ref19} opens up the possibility of success of the desired experiment in near future.
%\subsection{Subsection title}
\begin{table}[width=textwidth]
\centering
\caption{Some atomic properties and features of Ba$^{+}$ and Ra$^{+}$ related to APV experiment with single trapped and laser cooled ion. $^{\ast}$Approximate Calculation for $Q_{W}/N = 0.9$, $E'_{0} = 2 \times 10^{6}$ V/m, $f = 0.1$, $t = 24$ hrs. $^{\dag}$Calculated for $\lambda_{de} = 50$ nm.}
\label{tab2}       % Give a unique label
\begin{tabular}{lll}
\hline\noalign{\smallskip}
Atomic properties & $^{138}$Ba$^{+}$ & $^{226}$Ra$^{+}$  \\
\noalign{\smallskip}\hline\noalign{\smallskip}
Stability (neutral specis) & stable & $1620$ years ($T_{1/2}$) \\
$E1_{PNC}$ in $iea_{0}(-Q_{W}/N)\times 10^{-11}$ & $2.46$ \cite{Ref20} & $46.4$ \cite{Ref22} \\
PNC light-shift ($\Delta\omega^{PNC}_{m}/2\pi$) ($Hz$) & $0.3$ & $5.3$ \\
Coherence time $(\tau)$ (s) & $82$ & $0.6$ \\
$^{\ast}$Statistical uncertainty ($\frac{\delta E1_{PNC}}{E1_{PNC}}$) & $0.1 \%$ & $0.03 \%$ \\
$^{\dag}$Systematic uncertainty from $E'$ field (Eq. 5) & $1.2 \%$ & $7.1 \%$ \\
$^{\dag}$Systematic uncertainty from $E''$ field (Eq. 6) & $16 \%$ & $37 \%$ \\
Quenching rate of $S_{1/2} (m = 1/2)$\cite{Ref25} & 0.002 & 0.04 \\
Quenching rate of $D_{3/2} (m = 1/2)$\cite{Ref25} & 0.0033 & 3.36 \\
Quenching rate of $D_{3/2} (m = 3/2)$\cite{Ref25} & 0.0004 & 0.48 \\
\noalign{\smallskip}\hline
\end{tabular}
\end{table}
\section{Radium ion as favored candidate and challenges}
\label{sec:2}
Atomic parity violation effect scales little faster than $Z^{3}$ \cite{Ref2} for heavier element. In $^{226}$Ra$^{+}$ it is $20$ times larger as compared to $^{137}$Ba$^{+}$ \cite{Ref20} and $50$ times larger than that of atomic Cesium \cite{Ref21}. That is why at first sight $^{226}$Ra$^{+}$ is seemed to be a promising candidate though there are other advantages over Ba$^{+}$. The most recent calculation shows that the PNC amplitude present in $^{226}$Ra$^{+}$ is $46.4$ in the unit of $iea_{0}(-Q_{W}/N)\times 10^{-11}$ \cite{Ref22}, where $Q_{W}$ is the weak charge. In Table~\ref{tab2} the relevant atomic properties of $^{138}$Ba$^{+}$ and $^{226}$Ra$^{+}$ have been compared. The lasers required for $^{226}$Ra$^{+}$ are in visible and near infra-red region. Thus these lasers are available commercially as solid state diode lasers. Radium being a heavier element may be confined within smaller orbit than Barium as the Lamb - Dicke parameter is inversely proportional to the square root of mass of the ion. In addition, the known relative systematic uncertainties for Ra$^{+}$ are three times smaller as compared to Ba$^{+}$. The element has a large number of isotopes with significant stability, thus opening up the possibility of the experiment to extract the effect of nuclear structure in APV \cite{Ref23}. PNC amplitude $E1_{PNC}$ contains both nuclear spin dependent (NSD) and independent (NSI) parts \cite{Ref2}. From NSD part of $E1_{PNC}$, nuclear anapole moment can be measured \cite{Ref4}. However, in $S_{1/2} - D_{3/2}$ transition in Ra$^{+}$ or Ba$^{+}$, the contribution of NSD part is smaller by few orders than NSI part and hence determination of anapole moment is difficult. To avoid NSI part, a similar experiment explained above can be performed using $S_{1/2} - D_{5/2}$ transition of nuclear spin non-zero ($I \neq 0$) isotopes of Ra$^{+}$ or Ba$^{+}$. PNC allowed $S_{1/2} - D_{5/2}$ transition in these isotopes contains only NSD part and may lead to a direct measurement of nuclear anapole moment. $^{227}$Ra$^{+}$ is more favored candidate for an APV experiment than $^{137}$Ba$^{+}$ as PNC amplitude for $7S_{1/2} - 6D_{5/2}$ transition is $8$ times larger in this isotope \cite{Ref24}.

However, there are several disadvantages of choosing Ra$^{+}$ as a possible candidate. Trapping and cooling of Ra$^{+}$ has not been demonstrated so far. The lack of spectroscopic data on Ra$^{+}$ is also a problem. Theoretical calculation needs to be more accurate (below 1 \%) in order to compare with the experimentally obtained data. The atomic structure of Radium is not well studied. Coherence time is smaller for Ra$^{+}$ ($0.6 s$) which will reduce the signal to noise ratio. The systematic uncertainties originating from the determination of $E'$ and $E''$ at the ion position are too large for Ra$^{+}$ and demand some special experimental techniques to eliminate those. The production of various isotopes of Radium for the study of nuclear structure effects on APV demands well established facilities.

At the KVI, Groningen such a facility has been developed \cite{Ref22,Ref26} where some isotopes of Radium may be produced with an aim for performing APV experiment based on single trapped and laser cooled Ra$^{+}$. Atomic Ra has been successfully trapped in a MOT and laser cooled \cite{Ref27} at Argonne National Laboratory in search of permanent electric dipole moment (EDM) in atoms. Thus there is hope for details spectroscopic data on Ra to be available shortly which will lead towards the implementation of APV experiment on Ra$^{+}$.
\section{Present status in our group}
With an aim to perform high precision RF spectroscopy in search of APV, work has been started by our group RCAMOS at IACS. $^{138}$Ba$^{+}$ has been chosen initially as Barium is available commercially. A linear Paul trap has been designed. The repumping laser at $650$ nm has been frequency stabilized using Pound - Drever - Hall locking technique \cite{Ref28} and frequency doubling of $986$ nm laser \cite{Ref29} to produce cooling laser at $493$ nm is processing.

\begin{acknowledgements}
Part of this work is being supported by DST - SERC, India. P. Mandal and A. Sen are thankful to CSIR, India for entertaining fellowship during the research work. P. Mandal acknowledges the organizing committee of Laser 2009 Workshop, Poznan for sponsoring financial support to participate in the workshop.
\end{acknowledgements}


\begin{thebibliography}{spphys}
\bibitem{Ref1}% Format for Journal Reference
Fortson, N.: Phys. Rev. Lett. \textbf{70}, 2383 (1993)
\bibitem{Ref2}Bouchait,~M.~A., Bouchait, C.: Phys. Lett. B \textbf{48}, 111 (1974), Rep. Prog. Phys. \textbf{60}, 1351 (1997)
\bibitem{Ref3}Ginges, J. S. M., Flambaum, V. V.: Phys. Rep. \textbf{397}, 63 (2004), Marciano, W. J., J. L. Rosner, J. L.: Phys. Rev. Lett. \textbf{65}, 2963 (1990)
\bibitem{Ref4}Wood, C. S., Bennett, S. C., Cho, D., Masterson, B. P., Roberts, J. L., Tanner, C. E., Wieman, C. E.: Science \textbf{275}, 1759 (1997)
\bibitem{Ref5}Bennett, S. C., Wieman, C. E.: Phys. Rev. Lett. \textbf{82}, 2484 (1999)
\bibitem{Ref6}Meekohof, D. M., Vetter, P., Majumdar, P. K., Lamoreaux, S. K., Fortson, E. N.: Phys. Rev. Lett. \textbf{71}, 3442 (1993)
\bibitem{Ref7}Vetter, P. A., Meekhof, D. M., Majumder, P. K., Lamoreaux, S. K., Fortson, E. N.: Phys. Rev. Lett. \textbf{74}, 2658 (1995)
\bibitem{Ref8}Macpherson, M. J. D., Zetie, K. P., Warrington, R. B., Stacey, D. N., Hoare, J. P.: Phys. Rev. Lett. \textbf{67}, 2784 (1991)
\bibitem{Ref9}DeMille, David: Phys. Rev. Lett. \textbf{74}, 4165 (1995)
\bibitem{Ref10}Bouchait, M. A.: Phys. Rev. Lett. \textbf{100}, 123003 (2008)
\bibitem{Ref11}Koerber, T. W., Schacht, M., Nagourney, W., Fortson, E. N.: J. Phys. B: At. Mol. Opt. Phys. \textbf{36}, 637 (2003)
\bibitem{Ref12}Tsigutkin, K., Dounas-Frazer, D., Family, A., Stalnaker, J. E., Yashchuk, V. V., Budker, D.: Phys. Rev. Lett. \textbf{103}, 071601 (2009)
\bibitem{Ref13}Langacker, P., Luo, M., Mann, A. K.: Rev. Mod. Phys. \textbf{64}, 87 (1992)
\bibitem{Ref14}Majumdar, P. K., Tsai, L. L.: Phys. Rev. A \textbf{60}, 267 (1999)
\bibitem{Ref15}Leibfried, D., Blatt, R., Monroe, C., Wineland, D.: Rev. Mod. Phys. \textbf{75}, 281 (2003)
\bibitem{Ref16}Herskind, P. F., Dantan, A., Albert, M., Marler, J. P., Drewsen, M.: J. Phys. B: At. Mol. Opt. Phys. \textbf{42}, 154008 (2009)
\bibitem{Ref17}Sherman, J. A., Andalkar, A., Nagourney, W., Fortson, E. N.: Phys. Rev. A \textbf{78}, 052514 (2008)
\bibitem{Ref18}Sherman, J. A., Koerber, T. W., Markhotok, A., Nagourney, W., Fortson, E. N.: Phys. Rev. Lett. \textbf{94}, 243001 (2005)
\bibitem{Ref19}Koerber, T. W., Schacht, M. H., Hendrickson, K. R. G., Nagourney, W., Fortson, E. N.: Phys. Rev. Lett. \textbf{88}, 143002 (2002)
\bibitem{Ref20}Sahoo, B. K., Chaudhuri, R., Das, B. P., Mukherjee, D.: Phys. Rev. Lett. \textbf{96}, 163003 (2006)
\bibitem{Ref21}Dzuba, V. A., Flambaum, V. V., Ginges, J. S. M.: Phys. Rev. D \textbf{66}, 076013 (2002)
\bibitem{Ref22}Wansbeek, L. W., Sahoo, B. K., Timmermans, R. G. E., Jungmann, K., Das, B. P., Mukherjee, D.: Phys. Rev. A \textbf{78}, 050501(R) (2008)
\bibitem{Ref23}Fortson, E. N., Pang, Y., Wilets, L.: Phys. Rev. Lett. \textbf{65}, 2857 (1990)
\bibitem{Ref24}Geetha, K. P., Singh, A. D., Das, B. P.: Phys. Rev. A \textbf{58}, R16 (1998)
\bibitem{Ref25}Koerber, T. W., Thesis, Doctor of Philosophy, University of Washington (2003)
\bibitem{Ref26}Shidling, P. D., Giri, G.S., vanderHoek, D.J., Jungmann, K., Kruithof, W., Onderwater, C.J.G., Sohani, M., Versolato, O.O., Willmann, L., Wilschut, H.W.: Nucl. Instrum. Meth. A \textbf{606}, 305 (2009)
\bibitem{Ref27}Guest, J. R., Scielzo, N. D., Ahmad, I., Bailey, K., Greene, J. P., Holt, R. J., Lu, Z. T., O'Connor, T. P., Potterveld, D. H.: Phys. Rev. Lett. \textbf{98}, 093001 (2007)
\bibitem{Ref28}Drever, R. W. P., Hall, J. L., Kowalski, F. V., Appl. Phys. B \textbf{31}, 97 (1983)
\bibitem{Ref29}Raab, C., Bolle, J., Oberst, H., Eschner, J., Schmidt-Kaler, F., Blatt, R.: Appl. Phys. B \textbf{67}, 683 (1998)
\end{thebibliography}


\end{document}